\documentclass[aps,graphicx,preprint]{revtex4-1}
\usepackage{graphicx} \usepackage{amsmath} \usepackage{textcomp}
\draft % marks overfull lines with a black rule on the right

\begin{document}

%\preprint{}

\title{Large cooperativity and microkelvin cooling with a three-dimensional optomechanical cavity} %Title of paper



\author{Mingyun Yuan}
%\affiliation{Delft University of Technology}
\author{Vibhor Singh}
%\affiliation{Delft University of Technology}
\author{Yaroslav M. Blanter} \author{Gary A. Steele}
\email[]{g.a.steele@tudelft.nl}
%\thanks{}
%\altaffiliation{Delft University of Technology}
\affiliation{Kavli Institute of NanoScience, Delft University of
  Technology, PO Box 5046, 2600 GA, Delft, The Netherlands.}


\date{\today}

%\begin{abstract}
%\end{abstract}


\maketitle %\maketitle must follow title, authors, abstract and \pacs


\setlength{\parindent}{0cm}
%{\bf Abstract}

\textbf{In cavity optomechanics, light is used to control mechanical motion. A central goal of the field is achieving single-photon strong coupling, which would enable the creation of quantum superposition states of motion. Reaching this limit requires significant improvements in optomechanical coupling and cavity coherence. Here, we introduce a new optomechanical architecture consisting of a silicon-nitride membrane coupled to a three-dimensional superconducting microwave cavity. Exploiting their large quality-factors, we achieve an optomechanical cooperativity of 146,000 and perform sideband cooling of the kilohertz-frequency membrane motion to $34\pm5$ microkelvin, the lowest mechanical mode temperature reported to date. The achieved cooling is limited only by classical noise of the signal generator, and should extend deep into the ground state with superconducting filters. Our results suggest that this new realisation of optomechanics has the potential to reach the regimes of ultra-large cooperativity and single-photon strong coupling, opening up a new generation of experiments.}

%\vspace{1em}
\setlength{\parindent}{1cm}
%{\bf Introduction}

In recent years, cavity optomechanics has been used to realise a wide
range of exciting experiments with mechanical resonators, including
achieving exquisite measurement precision
\cite{teufel_nanomechanical_2009,anetsberger_measuring_2010,wilson_measurement_2014},
strong-coupling between the mechanical and cavity modes
\cite{Groblacher2009, Teufel2011a, Verhagen2012}, cooling to the
quantum ground state \cite{Teufel2011,Chan2011}, for microwave
amplification \cite{massel_microwave_2011,Singh2014}, to entangle
propagating microwave photons with mechanical motion
\cite{Palomaki08112013}, for microwave photon storage
\cite{palomaki_coherent_2013,zhou_slowing_2013}, to observe quantum
backaction noise \cite{Purdy2013a}, to generate squeezed light
\cite{squeezing2013,Purdy2013}, and to transduce photons between the
optical and microwave domains
\cite{bochmann_nanomechanical_2013,Andrews2014,Bagci2014}. These
successful experiments, performed in the regime of linear
optomechanics, were enabled by improvements of the coherence of the
optomechanical coupling between light and motion.

In linear optomechanics, the relevant figure of merit for the
optomechanical coupling is a parameter called
cooperativity. Cooperativity combines the optomechanical coupling rate
$g$, the cavity decay rate $\kappa$, and the mechanical decay rate
$\gamma_\text{m}$, into a dimensionless constant $C =
\frac{4g^2}{\kappa\gamma_\text{m}}$ that quantifies the optomechanical
system's efficiency in exchanging photons and phonons
\cite{RevModPhys.86.1391}. It is similar to the Purcell factor in
atomic physics (also often referred to as cooperativity) describing
the coupling between cavity fields and atoms.  An advantage of linear
optomechanics is that compared to the single-photon coupling rate
$g_0$, the multiphoton coupling rate $g$ is significantly enhanced,
given by $g = \sqrt{N_\text{d}}g_0$ where $N_\text{d}$ is the number of photons used to
drive the cavity.  The relevant limit for cooperativity in quantum
experiments is the so-called quantum coherent limit, in which the
cooperativity is larger than the number of equilibrium thermal quanta
in the mechanical resonator. In the quantum coherent limit, an
exchange of a photon and phonon occurs faster than the time it takes
for a phonon to leak into the mechanical resonator ground state from
the thermal bath. Achieving higher cooperativity that is deeper in the
quantum coherent limit in linear optomechanics would imply the ability
to cool closer to the quantum ground state and the preparation of
mechanical quantum states with high fidelity.

Beyond linear optomechanics, the field is striving to reach the limit
of single-photon strong coupling, in which interaction of light and
mechanical motion is coherent at the level of a single photon. In this
limit, the single-photon coupling rate $g_0$ exceeds both the cavity
decay rate $\kappa$ and the mechanical decay rate $\gamma_\text{m}$ ($g_0 >
\kappa,\gamma_\text{m}$).  If one could reach this limit, one could take
advantage of the intrinsic non-linearity of the optomechanical
coupling at the single-photon level to construct quantum
superpositions of mechanical motion using optomechanics with classical
light.  In current experiments, $g_0 \ll \kappa$, and existing
optomechanical implementations would require significant improvements
in $g_0$ and/or $\kappa$ to approach single-photon strong coupling.

Here, we present a new optomechanical architecture with large
optomechanical coupling that can potentially reach the single-photon
strong coupling limit. The design combines two highly coherent
elements that are applied for the first time in the microwave
optomechanics domain. The first element is a millimetre-sized,
nanometre-thick high-stress silicon nitride (SiN$_x$) membrane, a
technology that has demonstrated quality factors up to $1\times10^7$
at cryogenic temperatures \cite{Zwickl2008}. Such membranes have been
used extensively in the optical domain
\cite{Thompson2008,Purdy2013,Purdy2013a,Regal2014,Harris2014} and in
transducer applications \cite{Andrews2014,Bagci2014}, but have not yet
been explored as an element in a pure microwave optomechanical
system. The second optomechanical element is a three-dimensional (3D) microwave cavity,
recently popular in the superconducting qubit community for their
exceptional coherence times
\cite{Paik2011,kirchmair2013observation,Reagor2013}. By combining
these two highly coherent elements, we create a new optomechanical
platform with large cooperativity, demonstrate cooling to the lowest
mechanical mode temperature reported to date, and show that this new
system has the potential to scale to couplings significantly beyond
the state-of-the-art.

\begin{figure}
 \includegraphics[width=165mm]{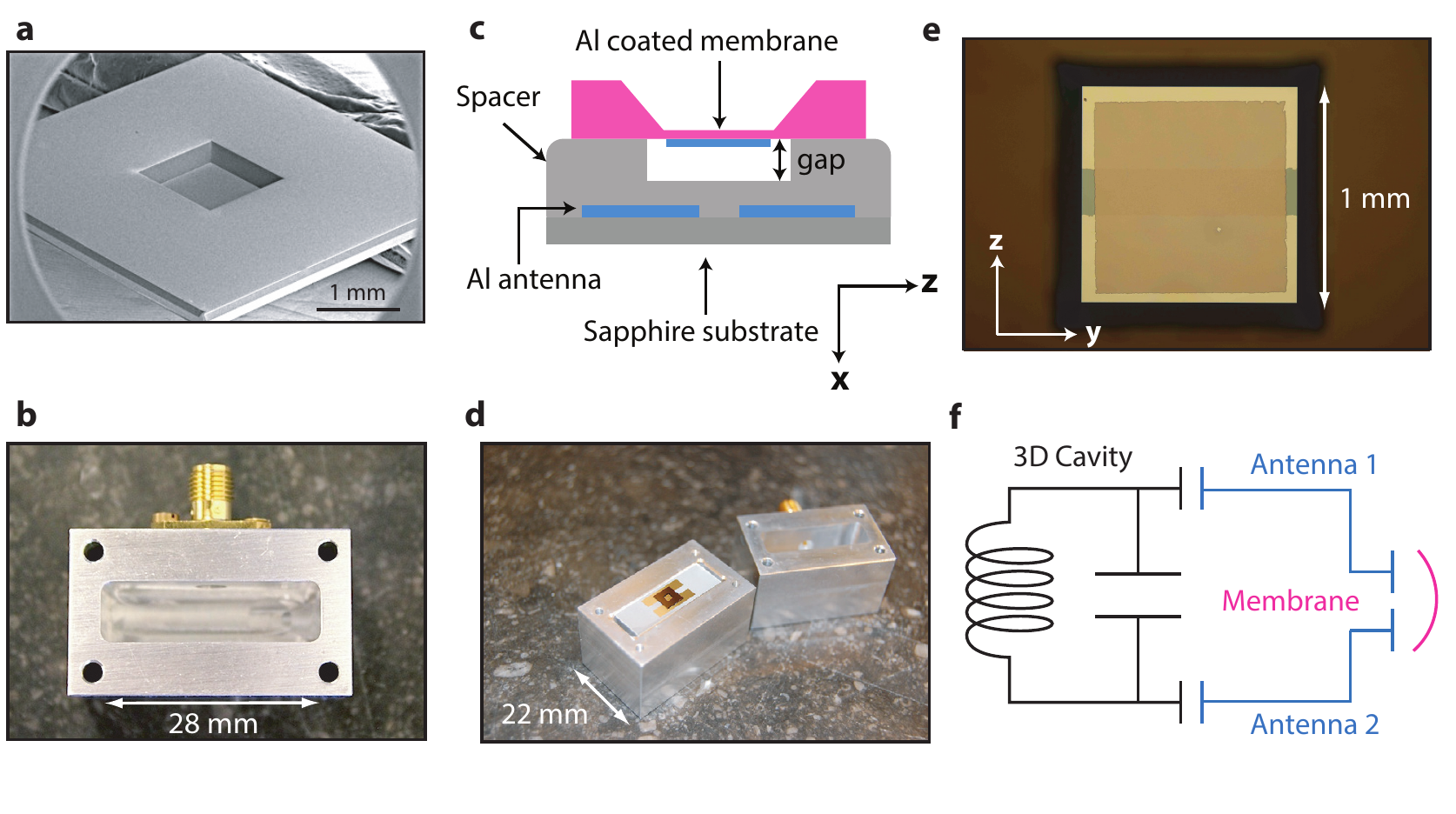}
 \caption{\label{dev}\textbf{Microwave optomechanics with a 3D
     superconducting cavity and a millimetre-sized membrane.}
   {\bf{a}}, Electron microscope image of a 50 nm thick SiN$_x$
   membrane. {\bf{b}}, One half of the Al cavity with an SMA connector
   for reflection measurements. The dimensions of the cavity are
   $28\times28\times8$ in millimetre. {\bf{c}}, Schematic showing the placement
   of the membrane over the antenna pads. The Al coating of the
   membrane forms a capacitor with the antennas below. {\bf{d}}, A
   complete assembly. The membrane is positioned in the centre of the
   cavity, supported by a sapphire substrate patterned with the Al
   antenna. {\bf{e}}, An optical microscope image looking from the top
   showing the Al-coated membrane and the underlying antenna
   pads. {\bf{f}}, Effective lumped element model of the cavity and
   membrane. By changing the effective shunt capacitance, the
   frequency of the 3D cavity is modulated by the mechanical motion of
   the membrane.}
 \end{figure}

\vspace{1em}
{\setlength{\parindent}{0cm}
{\bf Results}

{\bf Description and characterisation of the device.} Fig.~\ref{dev} illustrates the principle of this new 3D optomechanical
platform.  The mechanical resonator is made from high-stress SiN$_x$
membrane (Fig.~\ref{dev}a) that is metallised with an Al electrode.
The cavity itself is an aluminium box, Fig.~\ref{dev}b, in which
electromagnetic fields are confined by superconducting walls in all
three dimensions.  Coupling to the motion of the membrane is achieved
by using a flip-chip technique to place the membrane on top of antenna
electrodes on a separate substrate, Fig.~\ref{dev}c-e. Fig.~\ref{dev}f
shows an effective lumped-element circuit model of the assembled
cavity and membrane.  Optomechanical coupling results from the
modulation of the effective shunt capacitance when the membrane
displaces, changing the cavity frequency. From finite-element
simulations, we estimate a single photon coupling rate of
$g_0=(d\omega_0/dx) \cdot x_{\rm zpf} = 2\pi\times 0.36$~Hz, where $\omega_0$ is the cavity mode frequency, $x$ is the mechanical displacement of the membrane and $x_{\rm zpf}$ is the amplitude of its zero-point fluctuation. The
simplicity of the assembly of this 3D cavity architecture also makes
it attractive for implementing devices such as microwave-to-optical
transducers \cite{bochmann_nanomechanical_2013,Andrews2014,Bagci2014}
by potentially incorporating an optical fiber into the superconducting
box.}

Measurements are performed in a dilution refrigerator with a base
temperature of $T_\text{b}=13$~mK ({see Supplementary Figure 1 and Supplementary Note 1}). Fig.~\ref{char}a shows a measurement of
the reflection coefficient $|S_{11}|$ of the cavity. From a fit to the
data, we find a total linewidth of $\kappa=2\pi\times45.5$~kHz,
corresponding to a loaded quality factor of $Q_\text{L}=1.1\times10^5$. From
power dependence, we find a maximum intracavity photon occupation
$N_{\rm max} = 1.3\times10^8$ before the onset of a nonlinear
response. The ratio $\eta=\kappa_\text{e}/\kappa\approx0.48$ of the external
decay rate $\kappa_\text{e}$ and $\kappa$ indicates that the cavity is
slightly undercoupled ({see Supplementary Figure 2 and Supplementary Note 2}).

In Fig.~\ref{char}b, we characterise the mechanical response of the
membrane with the cavity using a resonant microwave tone injected at
$\omega_0$.  The thermomechanical motion of the membrane generates a
peak in the sideband power spectral density (PSD) $S(\omega)$ of the microwave field leaving the
cavity at a frequency offset $\Delta\omega=\omega_\text{m}$ from the carrier
signal, shown in Fig.~\ref{char}b.  We find a mechanical resonance
frequency of $\omega_\text{m} = 2\pi \times 123$~kHz, consistent with
expected fundamental mode frequency of the membrane.  A Lorentzian fit
yields a linewidth of $\gamma_\text{m}=2\pi\times3.5$~mHz, corresponding to an
ultra-high mechanical quality factor of $Q_\text{m}=3.5\times10^7$,
significantly higher than the typical $Q_\text{m} \lesssim 10^6$ of
membranes used in optomechanical and transducer experiments.

\vspace{1em}
\setlength{\parindent}{0cm}

{\bf Large cooperativity.} To quantify the strength of the optomechanical coupling, in
Fig.~\ref{coop} we measure the cooperativity $C$ between the
mechanical resonator and the cavity.  Together with initial phonon
occupancy of the mechanical resonator $n_\text{m}^\text{i}$, several criteria can be
conveniently expressed with $C$, such that for reaching the quantum
ground state of motion, $C+1 > n_\text{m}^\text{i}$, or for reaching the radiation
pressure shot noise limit, $C > n_m^\text{i}
(1+(\frac{\omega_\text{m}}{\kappa})^2)$.  To measure $C$, we use
optomechanically-induced transparency (OMIT)
\cite{Weis10122010,PhysRevA.81.041803,Singh2014}, which allows one to
directly determine the cooperativity with no free fit parameters.  In
OMIT, illustrated in Fig.~\ref{coop}a, the cavity is driven by a
strong drive tone ($\omega_\text{d}$) while a second weak probe ($\omega_\text{p}$)
is used to measure the cavity response. When driven on the red
sideband ($\omega_\text{d} = \omega_0 - \omega_\text{m}$), a transparency window
appears within the broad resonance dip of the cavity reflection
coefficient \cite{Singh2014}. In the limit $g\ll\kappa$, the linewidth
of the transparency window in $\vert S_{11}\vert^2$ is given by $\gamma_\text{m}(C+1)$ and the peak
value by $C/(C+1)$. Fig.~\ref{coop}b shows an example of an OMIT
measurement with drive-photon number $N_\text{d}=1.0\times10^8$: from the
broadened linewidth of the feature together with its near unity
transmission, we extract $C=94,500$. Fig.~\ref{coop}c shows the
extracted $C$ for different drive powers: at the maximum power
sustained by the cavity, we achieve $C_{\rm max}=1.46\times10^5$.

\vspace{1em}
{\bf Microkelvin cooling of the membrane resonator.} The large cooperativity of our optomechanical setup is, in principle,
capable of cooling deep into the quantum ground state of motion if
the mode is thermalised to the temperature of the fridge
($\frac{k_BT_\text{b}}{C_{\rm max}\hbar\omega_\text{m}} = 0.015$).  In order to
demonstrate the cooling of the resonator, in Fig.~\ref{cool} we use
the spectral density of the thermomechanical sideband to directly
observe the phonon occupation of the membrane.  While driving the
cavity on the red sideband, the output microwave PSD $S(\omega_0)$ at the cavity resonance frequency
$\omega_0$ is given by:
\begin{equation}\label{1}
\frac{S(\omega_0)}{\hbar\omega_0}=\frac{S_{vv}(\omega_0)}{\mathcal{G}\hbar\omega_0}=4\eta\frac{C}{(C+1)^2}\left(n_m^\text{i}+\frac{1}{2}\right)+n_{\rm
  add}
\end{equation}
where $S_{vv}(\omega_0)$ is the measured microwave PSD on the spectrum
analyser, $\mathcal{G}$ is the net gain of the signal path from the
output of the cavity to the input of the spectrum analyser, and
$n_\text{add}$ is the added photon noise quanta from the amplification
chain referenced to the output of the cavity.  By varying the bath
temperature $T_\text{b}$ and measuring $S_{vv}(\omega)$, we obtain an
absolute calibration of $\mathcal{G}$, $n_{\rm add}$, and $n_\text{m}^\text{i}$ ({details provided
in Supplementary Figure 3 and Supplementary Note 3, 4}).  At the base temperature of the refrigerator, we find that the
membrane is thermalised to $T_\text{m}^\text{i}\approx180$~mK, corresponding to
$n_\text{m}^\text{i}\approx3.06\times10^4$. From the thermal calibration, we also
extract $g_0=2\pi\times0.22$~Hz, in good agreement with simulations ({Supplementary Figure 4 and Supplementary Note 5}).

\setlength{\parindent}{1cm}

Fig.~\ref{cool}a shows $S(\omega)$ for different cooperativities $C$ of the
cooling tone. As $C$ increases, $S(\omega_0)$ from the thermomechanical noise peak
decreases, indicating that the mode temperature of the mechanical
resonator is reduced. At larger cooling powers, however, although
$S(\omega_0)$ continues to drop, the noise floor of $S(\omega)$
outside the mechanical bandwidth begins to increase, and the peak in
the PSD becomes a dip (bottom two panels of Fig.~\ref{cool}a). The
increase in the noise floor is an indication of noise fluctuations of the cavity
field \cite{Teufel2011}. Due to correlations between the fluctuations
of the cavity and of the mechanical resonator, the spectrum shows a
suppression of the total PSD at the cavity frequency
\cite{PhysRevLett.99.017201}.

In order to extract the mechanical occupation factor in the presence
of cavity noise, it is no longer sufficient to look only at
$S(\omega_0)$.  In particular, with sufficiently large $C$ (and in the
absence of other sources of heating that would increase $n_\text{m}^\text{i}$ such
as losses in the superconducting film or in the dielectric membrane or
substrate), $S(\omega_0)$ drops to
the amplifier noise floor independent of the amount of cavity noise
(Eq.~\ref{1}). This does not, however, imply that the mechanical mode
is at zero temperature: due to the hybridisation of the mechanical and
optical fields, the final mechanical mode occupation in the presence
of cavity noise is given by ({Supplementary Note 6}):
\begin{equation}
  n_\text{m} = \frac{1}{(C+1)} n_\text{m}^\text{i} + \frac{C}{(C+1)} n_\text{c}
\end{equation}
where $n_\text{c}$ is the cavity noise power measured in energy quanta.
In order to find the final occupation $n_\text{m}$, one must also determine
$n_\text{c}$.  In the limit $g\ll\kappa$, cavity noise appears as an increase
in the noise floor outside of the mechanical sideband (green dashed
line in Fig.~\ref{cool}a) with a spectral density ({see Supplementary Note 6 for the more general equation of $S(\omega)$}):
\begin{equation}
  \frac{S(\omega_0 + \delta \omega)}{\hbar \omega} = 4 \eta \left( n_\text{c}
  + \frac{1}{2} \right) + n_{\rm add}
\end{equation}
where $(C+1)\gamma_\text{m} \ll \delta\omega \ll \kappa$. From this
expression, one can extract the cavity occupation $n_\text{c}$ and
consequently $n_\text{m}$ for all powers, shown in Fig.~\ref{cool}b. As $C$
is increased, we find that $n_\text{m}$ drops initially to a value of $5.2\pm0.7$,
corresponding to a mode temperature of $34\pm5$~$\mu$K, beyond which the
mechanical occupation begins to increase sharply due to heating from
cavity noise. The temperature reported here is roughly a
factor of two lower than recent experiments in the optical domain with
silicon nitride membranes \cite{Regal2014, Harris2014} due to the very low
frequency of our membrane.

Although the large cooperativity in our experiment should allow us to
cool the membrane to $n_\text{m} = 0.2$ given the initial thermal occupation,
in practice we are limited to $n_\text{m} = 5.2\pm0.7$ by cavity noise. In order to
cool to lower occupation in future experiments, it is important to
identify the source of this cavity noise. The green line in Fig. 4b
shows the expected $n_\text{c}$ due to the carrier sideband noise of our
microwave signal generator ({Supplementary Note 7}). The good agreement with the observed
cavity noise data suggests that our final occupation is limited by the
spectral purity of the microwave tone used for the sideband cooling.

\vspace{1em}
\setlength{\parindent}{0cm}
{\bf Discussion}

Having demonstrated the lowest temperature $T_\text{m} = 34\ \mu$K reported
to date for a mechanical resonator, we analyse the potential of this
new implementation to reach the deep quantum coherent coupling limit.
In the current experiment, the cooling is limited by the classical
sideband noise of the signal generator.  Removing this noise with a
tunable superconducting cavity with a linewidth of 10~kHz would
already provide sufficient suppression to cool to a final occupation
of 0.2.  A second approach would be to increase $g_0$ by shrinking the
capacitor gap: reducing the membrane-antenna gap from 3~$\mu$m to 100~nm would
increase $g_0$ by a factor of $10^3$ and $C_{\rm max}$ by $10^6$.  A
third approach is to improve the cavity linewidth \cite{Reagor2013},
also yielding higher cooperativity at lower photon numbers. Finally, combining a smaller
gap ($g_0\sim 300$ Hz) with a better cavity ($\kappa \sim 10$ Hz)
could yield ultra-high cooperativities $C > 10^{12}$. Such a fully
optimised design would also achieve single-photon strong coupling
($g_0 > \kappa,\gamma_\text{m}$)
\cite{rabl2011photon,PhysRevLett.107.063602}, enabling preparation and
detection of non-classical states of motion such as Fock states or
Schr\"odinger cat states with optomechanics.

\setlength{\parindent}{1cm}

In conclusion, we have developed a novel optomechanical system
coupling the motion of a millimetre-sized membrane to a 3D microwave
cavity. Exploiting the high coherence of the membrane and of the
cavity, we achieve a cooperativity of $C > 1.4\times10^5$ and perform
sideband cooling of the millimetre-size membrane to 34~$\mu$K,
corresponding to a thermal occupation $n_\text{m} = 5.2$. The scaling of this
3D optomechanical system offers the possibility to reach
optomechanical couplings far beyond the state-of-the-art, potentially
entering the single-photon strong coupling regime in which a new
generation of quantum experiments with mechanical objects would become
possible.

\setlength{\parindent}{0cm}

\vspace{2em}
{\bf Acknowledgments} We would like to thank S. J. Bosman, S. Yanai,
S. Gr\"oblacher, L. DiCarlo, D. Rist\`e and R. Hoogerheide for
discussions and support. Fabrication is carried out in Kavli Nanolab
and this project is supported by the Stichting voor Fundamenteel
Onderzoek der Materie (FOM).

\vspace{2em}
{\bf{Author Contribution}} G. A. S. conceived the device
and supervised the project. M. Y. prepared the devices. M. Y. and
V. S. set up the experiment and performed the measurements. M. Y.,
V. S. and G. A. S. analysed the data. M. Y. and Y. M. B. performed the
theoretical calculation. All authors contributed to writing of the
manuscript.

\vspace{2em}
{\bf Methods}

{\bf Device preparation.} We use commercial SiN$_x$ membranes manufactured by Norcada. The
membranes have the dimensions of
$50~\text{nm}\times1~\text{mm}\times1~\text{mm}$, and are supplied
with a 5 mm $\times$ 5 mm Si frame. We deposit a 20~nm thick film of
Al on top of the membrane without covering the clamping edges by using
a physical mask.  On a separate sapphire substrate, we pattern two Al
antenna pads with 80~nm of Al followed by the deposition of SiN$_x$
spacer layer. To ensure the membrane does not come into contact with
the substrate, a recess of 100~nm is etched. The metallised SiN$_x$
membrane is then placed on top of the antenna pads to form a capacitor
using a vacuum pick-and-place technique. A single drop of 0.1~$\mu$l
of 2-part epoxy is applied on the substrate to attach one corner of
the membrane's Si frame to the substrate.  Using the depth of focus to
locate the vertical position of the membrane and of the bottom antenna
pads while looking through the membrane with an optical microscope, we
estimate the gap to be approximately $3~\mu$m. The gap is much larger than that designed in the spacer, most likely due to contamination from dust in the large contact area between the Si frame of the membrane and the substrate. The 3D microwave
cavity is formed by closing two halves of a machined block made out of
6061-aluminium.  The inner surface of the cavity is polished using a
polishing paste, but is not chemically etched. The cavity is assembled
by screwing the two halves of the cavity together with no sealing
mechanism.  The dimension of the whole cavity is
28~mm({\bf x})$\times$28~mm({\bf y})$\times$8~mm({\bf z}), with rounded corners of
radius 1~mm in the {\bf y-z} plane. 

\vspace{1em}
{\bf Measurement.} The bare frequency of the cavity without
the antenna substrate and membrane is 7.4~GHz. The membrane attached
to the antenna is placed at the centre of the cavity, coupling to the
TE$_{110}$ mode.  We estimate the total mass of the membrane including
the Al layer to be $m=200$~ng, corresponding to an estimated quantum
zero-point fluctuation of $x_{\rm zpf}=0.6$~fm. Measurements are performed
in a cryogen free dilution refrigerator. To minimise cavity
instability during the measurements, the sample is mounted on a
mass-spring vibration isolation stage on the mixing chamber with
resonance frequency of $\sim 1$ Hz. During the cooling measurements,
the pulse tube is temporarily turned off to minimise vibrations of the
sample and the cables. Measurements are started 1.5 minutes after the
switch-off and performed in a $\sim 2$ minute window before the
temperature of the mixing chamber begins to increase. %Detailed measurement setup is explained in Supplementary Figure and Supplementary Note  .


\newpage

 \begin{figure}
 \includegraphics[width=88mm]{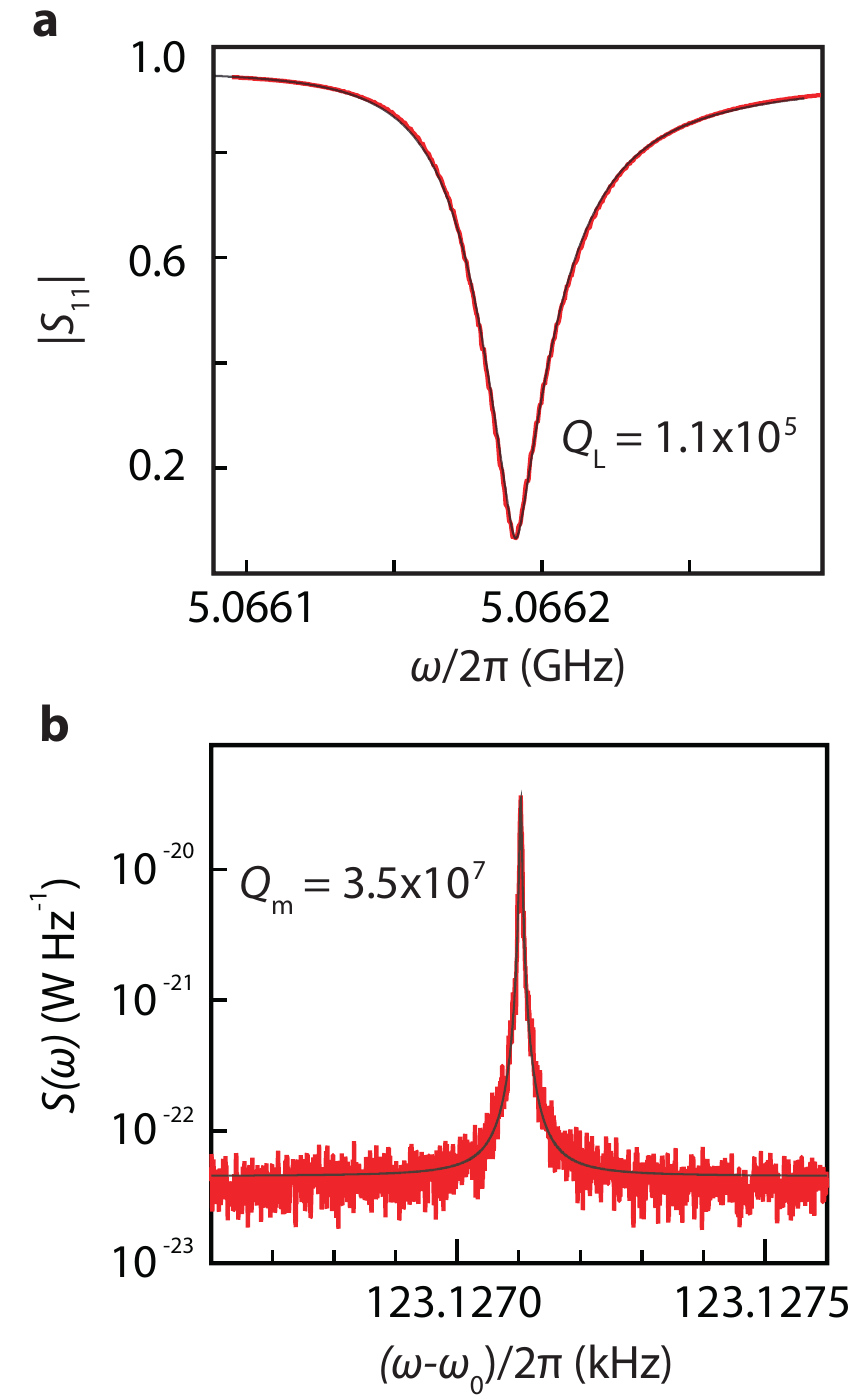}
 \caption{\label{char}\textbf{High quality-factors of the microwave
     cavity and mechanical resonator.}  {\bf{a}}, Reflection
   coefficient $|S_{11}|$ measurement of the cavity. The loaded
   quality factor is $Q_\text{L}=1.1\times10^5$, corresponding to a total
   decay rate $\kappa=2\pi\times45.5$~kHz. Internal dissipation rate is
   $\kappa_0=2\pi\times23.9$~kHz, with $Q_0~=~2\times10^5$. {\bf{b}}, Power
   spectral density of the thermomechanical motion of the fundamental mode of
   the membrane $\omega_\text{m}\approx2\pi\times123$~kHz with linewdith
   $\gamma_\text{m}=2\pi\times3.5$~mHz, corresponding to $Q_\text{m} =
   3.5\times10^7$. Red: data, Grey: fit.}
 \end{figure}

\begin{figure}
\includegraphics[width=165mm]{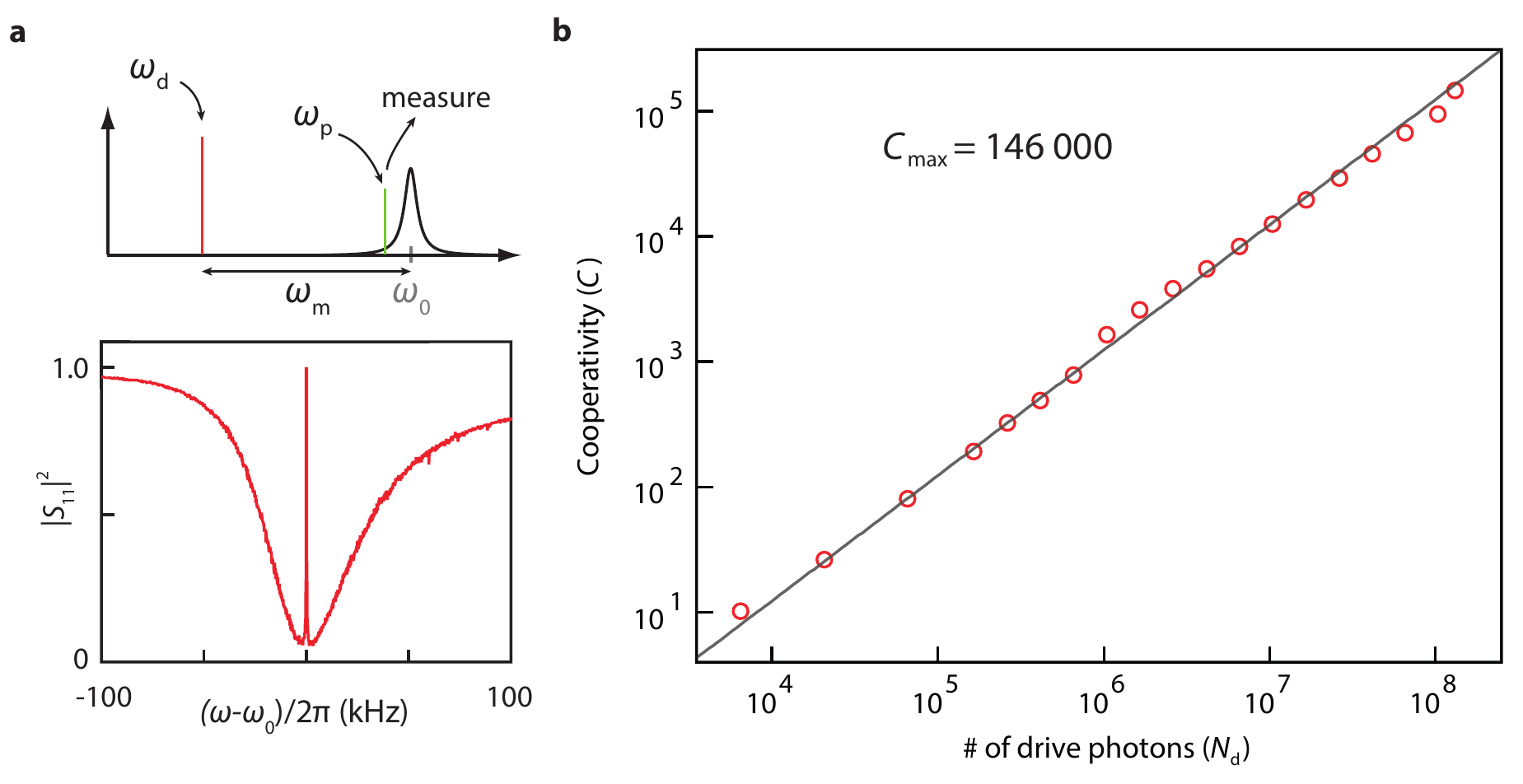}
\caption{\label{coop}\textbf{Large cooperativity measured with
    optomechanically-induced transparency (OMIT).} {\bf{a}},
  Illustration showing the OMIT measurement scheme. The cavity
  reflection coefficient $S_{11}$ is measured with a weak probe tone
  ($\omega_\text{p}$) while driving the cavity with a second strong tone near
  the red sideband ($\omega_\text{d} = \omega_0-\omega_\text{m}$). A window of
  transparency appears in the cavity resonance (lower panel) with a
  width set by the mechanical linewdith. {\bf{b}}, Extracted
  cooperativity $C$ vs. driving photon number $N_\text{d}$. Grey line:
  expected linear scaling $C = \frac{4g_0^2 N_\text{d}}{\gamma_\text{m} \kappa}$. At
  maximum $N_\text{d}$, the cooperativity reaches $C_{\rm
    max}=1.46\times10^5$, corresponding to $g_{\rm max}=
  2\pi\times2.5$~kHz.}
\end{figure}

\begin{figure}
 \includegraphics[width=165mm]{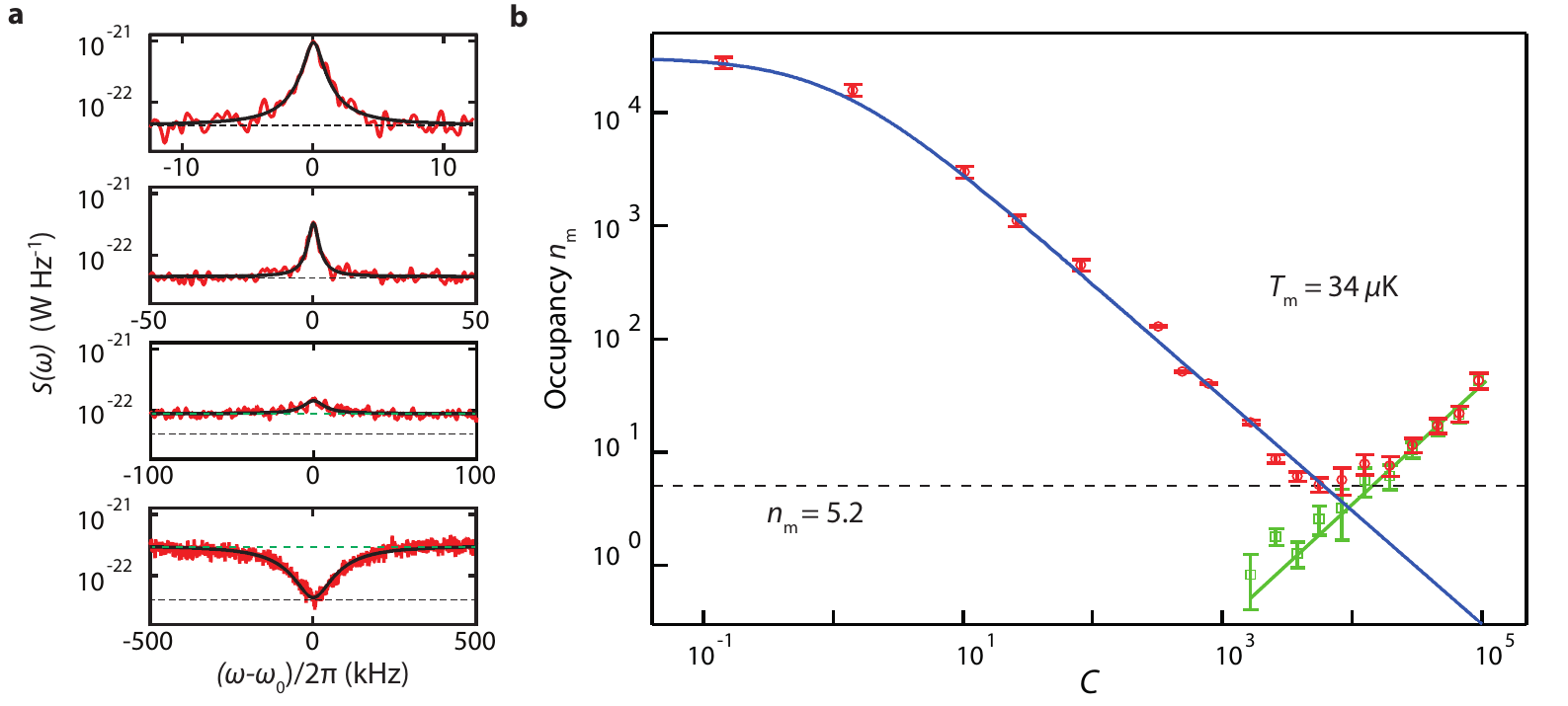}
 \caption{\label{cool}\textbf{Microkelvin cooling of a millimetre-sized mechanical
     resonator.} {\bf{a}}, Measured microwave PSD $S(\omega)$ with
   red-sideband driving (top to bottom: $C=324$, 785, 3810,
   94500). Black dashed line: noise floor $n_\text{add}$ of the amplifier
   chain. As the cooperativity $C$ of the cooling tone is increased,
   $S(\omega_0)$ is reduced and the mechanical linewdith is broadened.
   Initially, the thermal noise of the membrane appears as a peak with
   decreasing height.  At higher $C$ (lower two panels), the noise floor
   outside of the mechanical linewidth (green dashed line) begins to
   increase above $n_\text{add}$, indicating the presence of cavity noise.
   In the lowest panel, the output noise of the cavity is suppressed
   at $\omega_0$ due to correlations of the mechanical and cavity
   fluctuations.  From the thermal peak and the cavity noise background level (green dashed
   line), we extract both $n_\text{m}$ and $n_\text{c}$ for all cooling powers,
   shown in {\bf b}.  As a function of $C$, the mechanical occupancy
   $n_\text{m}$ (red circle) drops, closely following $\frac{n_m^\text{i}}{C+1}$
   (blue line) until the onset of cavity noise $n_\text{c}$ (green squares)
   limits the final occupancy to a minimum of $n_\text{m}$~=~5.2,
   corresponding to $T_{\rm m} = 34\ \mu$K. The error bars indicate the uncertainty of the data points and are calculated with the errors from the Lorentzian fit. The green line in {\bf b}
   shows the expected $n_\text{c}$ from the measured carrier sideband noise of the
   microwave generator.}
 \end{figure}

\end{document}

\renewcommand{\figurename}{\bf Supplementary Figure}

%\preprint{}

%\title{Supplementary information:\\
%Large cooperativity and microkelvin cooling with a 3D
%optomechanical cavity} %Title of paper



%\author{Mingyun Yuan}
%\affiliation{Delft University of Technology}
%\author{Vibhor Singh}
%\affiliation{Delft University of Technology}
%\author{Yaroslav M. Blanter}
%\author{Gary A. Steele}
%\thanks{}
%\altaffiliation{Delft University of Technology}
%\affiliation{Kavli Institute of NanoScience, Delft University of Technology, PO Box 5046, 2600 GA, Delft, The Netherlands.}

%\date{\today}
%\maketitle
\begin{figure}
 \includegraphics[width=120mm]{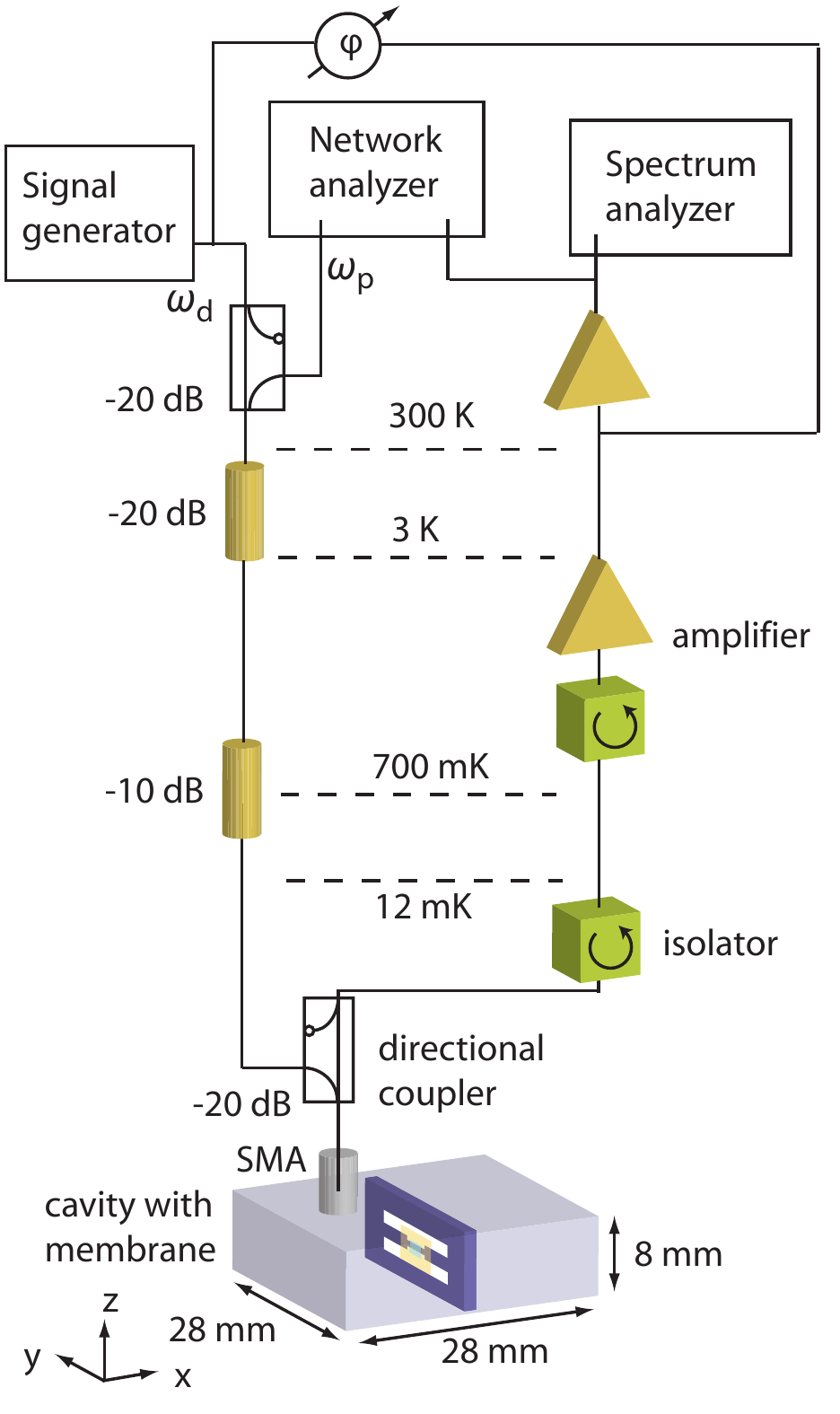}
 \caption{\label{setup} {\bf Experimental setup for reflection measurements.}}
 \end{figure}
\clearpage

\begin{figure}
 \includegraphics[width=120mm]{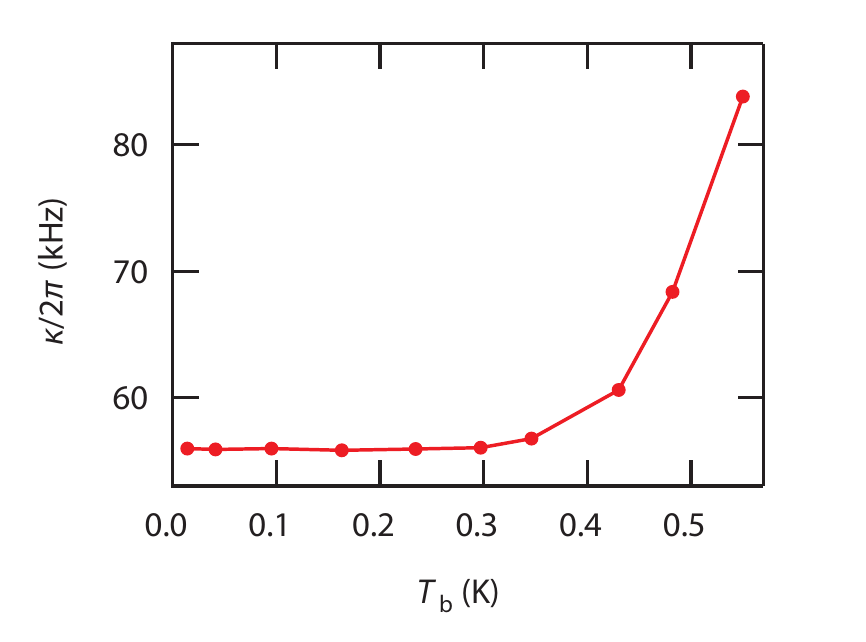}
 \caption{\label{Tdep}Cavity decay rate $\kappa$ as a function of cryostat base temperature $T_\text{b}$. \rm The line width of the cavity resonance starts to increase above 0.34~K.}
 \end{figure}

\begin{figure}
 \includegraphics[width=100mm]{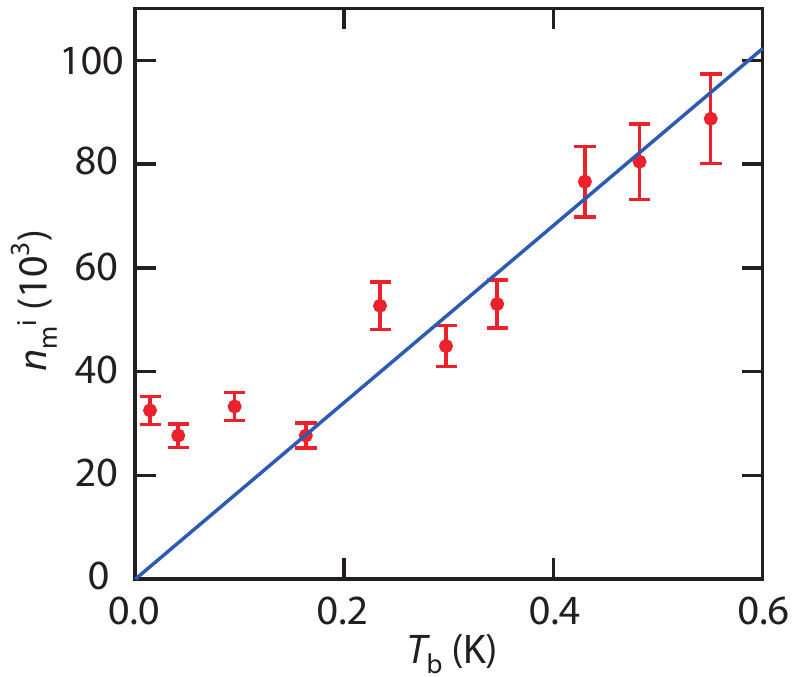}
 \caption{\label{Tcalib}Calibration of the initial thermal occupancy $n_\text{m}^\text{i}$ by varying the refrigerator temperature $T_\text{b}$. \rm The mode temperature of the membrane is thermalised at base temperature to $T_\text{m}^\text{i}\approx180$~mK, corresponding to $n_\text{m}^\text{i}\approx3.06\times10^4$. The error bars indicate the uncertainty of the data points, combining the fluctuations due to the temperature and the signal level.}
 \end{figure}
\clearpage

 \begin{figure}
 \includegraphics[width=120mm]{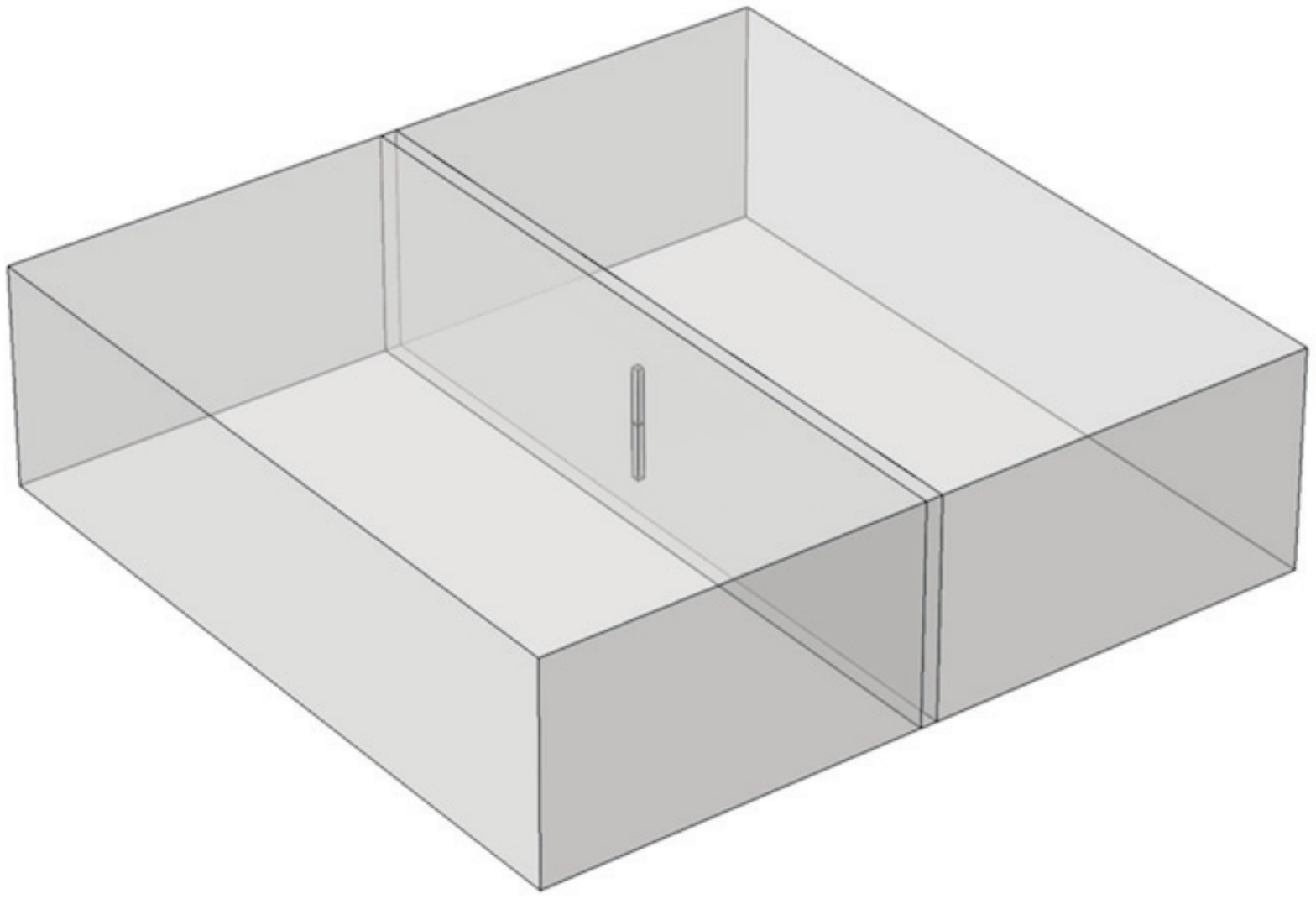}
 \caption{\label{geo}COMSOL model showing a 3D cavity embedding an antenna.}
 \end{figure}
\clearpage

\setlength{\parindent}{0cm}
{\bf \large Supplementary Note 1: Experimental setup}
\vspace{1em}

The sample is cooled down in a cryogen-free dilution refrigerator to a
base temperature of 12 to 13~mK. Input microwave signals are
attenuated at each thermal stage as illustrated by Supplementary Figure~\ref{setup}.
At base temperature, the signal is injected into the cavity via the
coupling port of a 20~dB directional coupler. The reflected power
collected at the output port is amplified by a cold amplifier at the
3~K plate and subsequent amplifiers at room temperature and then
detected by a network analyser or a spectrum analyser. Circulators are
placed between the sample and the cold amplifier to isolate the sample
from noise generated by the amplifier. A cancelling tone is applied
using a directional coupler, a phase shifter and a continuous
attenuator in order to reduce the amplitude of the strong drive tone
before sending it to the room temperature amplifier. For the optomechanically-induced transparency (OMIT)
experiment, a network analyser is used to generate the weak probe tone
$\omega_\text{p}$ and a separate phase locked signal generator provides the
strong drive tone $\omega_\text{d}$.\\

{\bf \large Supplementary Note 2: Cavity characterisation}
\vspace{1em}

The cavity response $\left\vert S_{11}\right\vert$ in Fig. 2a in the main text is fitted with the following equation, taking into consideration the finite isolation of the directional coupler \cite{Singh2014}:
\begin{equation}\label{S11}
\left\vert S_{11}(\omega)\right\vert=\left\vert\alpha e^{i\phi}+(1-\alpha)(1-\frac{\kappa_\text{e}}{i(\omega-\omega_0)+\frac{\kappa}{2}})\right\vert,
\end{equation}
where $\alpha$ is the isolation of the directional coupler.

\setlength{\parindent}{1cm}
Due to the low superconducting transition temperature of aluminium $T_c\sim1.2$~K, the
internal cavity quality factor is significantly reduced as the cryostat
temperature $T_\text{b}$ is raised above $\sim$0.34~K, and the line width
$\kappa$ is broadened, as shown in Supplementary Figure~\ref{Tdep}. This effect needs to be taken into account in the
thermal calibration. As a function of temperature, we find that
$\kappa_\text{e}$ remains constant.\\

\setlength{\parindent}{0cm}
{\bf \large Supplementary Note 3: Calibration of the mode temperature}
\vspace{1em}

A thermal calibration is carried out to determine the initial
(i.e. without cooling) phonon occupancy
$n_\text{m}^\text{i}=\frac{1}{\text{exp}{\frac{\hbar\omega_\text{m}}{k_\text{B} T_\text{m}^\text{i}}}-1}$ of the
mechanical resonator, $k_\text{B}$ being the Boltzmann constant and $T_\text{m}^\text{i}$
the initial mode temperature. We send in a carrier signal at
$\omega_0$ to avoid any backaction effects and measure the total power
of its mechanical sideband at different cryostat temperature $T_\text{b}$. The
total power generated by the thermal motion of the membrane is then
used to extract the mechanical mode temperature $T_\text{m}^\text{i}$. We use a
resolution bandwidth (RBW) of 1~Hz. In the limit of RBW$\gg\gamma_\text{m}$,
the total thermal power can be directly read out from the peak
height. For $T_\text{b}>0.34$~K, $\omega_0$ and $\kappa$ changes significantly
due to the low critical temperature of Al, which has to be
taken into account. We plot the converted $n_\text{m}^\text{i}$ as a function of
refrigerator temperature $T_\text{b}$ in Supplementary Figure~\ref{Tcalib}. From the fitting
line that passes the original point, at the base temperature the
membrane is thermalised to $T_\text{m}^\text{i}\approx180$~mK and
$n_\text{m}^\text{i}\approx3.06\times10^4$. The mode temperature is significantly
higher than the base temperature of the refrigerator, which is
probably an inevitable challenge one is faced with when working with a
low-frequency ($<1$~MHz) mechanical resonator.\\

{\bf \large Supplementary Note 4: Power calibration}
\vspace{1em}

%\section{Equations for photon numbers and sideband power}

The photon number in the cavity corresponding to the input power at the cavity $P_\text{in}$ at frequency $\omega$ is calculated with
\begin{equation}\label{nd}
N=\frac{P_\text{in}}{\hbar\omega}\cdot\frac{\kappa_\text{e}}{(\frac{\kappa}{2})^2+(\omega-\omega_0)^2}.
\end{equation}
While driving the cavity on resonance, the sideband power due to the thermal motion is expressed as
\begin{equation}\label{Pside}
\begin{aligned}
P_\text{side}=&P_\text{in}\left(\frac{g_0}{x_\text{zpf}}\right)^2\cdot\frac{k_\text{B}T}{m\omega_\text{m}^2}\cdot\frac{(\frac{\kappa_\text{e}}{2})^2}{(\frac{\kappa}{2})^2+(\omega-\omega_\text{r,b})^2}\frac{1}{(\frac{\kappa}{2})^2+(\omega-\omega_0)^2}\\
=&P_\text{in}\cdot g_0^2\cdot2n_\text{m}\cdot\frac{(\frac{\kappa_\text{e}}{2})^2}{(\frac{\kappa}{2})^2+(\omega-\omega_\text{r,b})^2}\frac{1}{(\frac{\kappa}{2})^2+(\omega-\omega_0)^2},
\end{aligned}
\end{equation}
where $g_0$ is again the single-photon coupling rate, $k_\text{B}$ the Boltzmann constant and $n_\text{m}$ the mechanical occupancy. The sideband frequency of interest is either $\omega_\text{r}$ or $\omega_\text{b}$.
\setlength{\parindent}{1cm}

The loss of the input line $\mathcal{L}$ and the gain of the output line $\mathcal{G}$ are calibrated as the following:
while driving the cavity on resonance ($\omega=\omega_0$) with power $P_\text{in}'$ from the signal generator, the thermomechanical power measured by the spectrum analyser at the mechanical sidebands can be written as:
\begin{equation}
\begin{aligned}
P_\text{side}=&P_\text{in}'\cdot g_0^2\cdot2n_\text{m}\left(\frac{\kappa_\text{e}}{\kappa}\right)^2\frac{1}{(\frac{\kappa}{2})^2+\omega_\text{m}^2}\cdot \mathcal{L} \cdot \mathcal{G}\\
=&P_\text{in}'\cdot g_0^2\cdot\beta.
\end{aligned}
\end{equation}
The combined value of $\mathcal{L} \cdot \mathcal{G}$ is determined by the network analyser. Thermal occupation can be measured by sweeping the bath temperature as described in the previous section. All the parameters being known in $\beta$, single-photon coupling strength $g_0$ can be extracted. This is equivalent to the single-photon coupling strength calibration the frequency modulation (FM) technique described in \cite{Gorodetsky:10}, where the FM peak of the carrier wave provides a side-by-side reference of $\mathcal{L} \cdot \mathcal{G}$. Using this approach, we measure the single-photon coupling strength to be $g_0=0.22$~Hz. This further allows us to calibrate the corresponding photon number $N$ by using the experimentally obtained photon-enhanced $g=g_0\sqrt{N}$ (from cooperativity measurements).
Subsequently, by rewriting Supplementary Equation~\ref{nd} as
\begin{equation}
N=\frac{P_\text{in}'}{\hbar\omega}\cdot\frac{\kappa_\text{e}}{(\frac{\kappa}{2})^2+(\omega-\omega_0)^2}\cdot \mathcal{L},
\end{equation}
we calibrate the total input attenuation $\mathcal{L}=70$~dB and correspondingly the output gain $\mathcal{G}=73.5$~dB. The added noise of the output chain is calibrated to be $n_\text{add}=12$ corresponding a noise temperature of approximately~2.6~K.\\

{\setlength{\parindent}{0cm}
{\bf \large Supplementary Note 5: COMSOL RF simulations to estimate $g_0$}
\vspace{1em}

To estimate the single-photon coupling rate $g_0$ of the 3D-cavity-membrane system, we model the problem with a ``variable capacitor in cavity" structure and numerically calculate the cavity frequency $\omega_0$ with COMSOL. The following relation is used to derive $g_0$:
\begin{equation}
g_0=\frac{\delta\omega_0}{\delta C_1}\cdot\delta C_2
\end{equation}
where $\delta C_1$ is the change in capacitance by varying the gap between the two antenna rods, $\delta\omega_0$ the subsequent frequency shift and $\delta C_2$ the change in capacitance due to zero-point fluctuation $x_\text{zpf}$ of the membrane. $\delta C_2$ can be expressed as $\delta C_2=\frac{\epsilon_0 A_2}{d^2} x_\text{zpf}$, where $\epsilon_0$ is the free space permittivity, $A_2$ the area of the membrane and $d$ the gap between the membrane and the antenna.}

The geometry of the COMSOL model is shown in Supplementary Figure~\ref{geo}. The dimension of the 3D cavity is 28~mm$\times$28~mm$\times$8~mm. There is a 0.6~mm thick sapphire substrate at the centre of the cavity. The antenna is simplified to two Al rods with a cross section of 0.25~mm$\times$0.25~mm, the length of the antenna being 4~mm and their gap is varied between 20 and 30~$\mu$m, changing the parallel-plate capacitance between them. Although this geometry is different from the actual geometry of the antenna, it results in comparable frequency pull of $\omega_0$, indicating that the capacitance participation ratio in the equivalent lumped-element circuit is similar. For a membrane of 1~mm$\times$1~mm, a gap of $d=3$~$\mu$m and $x_\text{zpf}=0.6$~fm the resulting coupling strength is $g_0\approx0.36$~Hz. Since $g_0$ is inversely proportional to the square of $d$, if $d$ can be reduced to 30~nm, $g_0$ can be increased to 3.6~kHz.\\

{\setlength{\parindent}{0cm}
{\bf \large Supplementary Note 6: Quantum noise of optomechanical cooling}
\vspace{1em}

The linearised Hamiltonian of the optomechanical system in a frame rotating at the drive tone frequency $\omega_\text{d}$ can be written as \cite{RevModPhys.86.1391, Marquardt2007, PhysRevLett.101.263602, Rocheleau2010, Teufel2011}:
\begin{equation}
\hat{H}=-\hbar\Delta\hat{a}^\dagger\hat{a}+\hbar\omega_\text{m}\hat{b}^\dagger\hat{b}-\hbar g(\hat{a}^\dagger+\hat{a})(\hat{b}^\dagger+\hat{b}),
\end{equation}
where $\Delta=\omega_\text{d}-\omega_0$, $\hat{a}^\dagger$($\hat{a}$) the creation (annihilation) operator for the cavity field variation and $\hat{b}^\dagger$($\hat{b}$) the creation (annihilation) operator for the mechanical mode. The optomechanical coupling $g=g_0\sqrt{N}$ is enhanced by the number of photons $N$, $g_0$ being the single-photon coupling rate.}

We obtain the Heisenberg-Langevin equations
\begin{align*}
 \dot{\hat{a}}(t) =& \left(i\Delta-\frac{\kappa}{2}\right)\hat{a}(t) + ig(\hat{b}(t)+\hat{b}^{\dagger}(t)) + \sum_{j=\text{e},0}\sqrt{\kappa_j}\hat{\xi}_j(t)\\
 \dot{\hat{b}}(t) =& \left(-i\omega_\text{m}-\frac{\gamma_\text{m}}{2}\right)\hat{b}(t) + ig(\hat{a}(t)+\hat{a}^{\dagger}(t)) + \sqrt{\gamma_\text{m}}\hat{\xi}_\text{m}(t)
\end{align*}
and their Hermitian conjugates
\begin{align*}
 \dot{\hat{a}}^{\dagger}(t) =& \left(-i\Delta-\frac{\kappa}{2}\right)\hat{a}^{\dagger}(t) - ig(\hat{b}(t)+\hat{b}^{\dagger}(t)) +  \sum_{j=\text{e},0}\sqrt{\kappa_j}\hat{\xi}_j(t)\\
 \dot{\hat{b}}^{\dagger}(t) =& \left(i\omega_\text{m}-\frac{\gamma_\text{m}}{2}\right)\hat{b}^{\dagger}(t) - ig(\hat{a}(t)+\hat{a}^{\dagger}(t)) + \sqrt{\gamma_\text{m}}\hat{\xi}_\text{m}(t),
\end{align*}
where $\kappa$ and $\gamma_\text{m}$ are the decay rates of the cavity and the mechanical resonator respectively. The noise operators satisfy the following relationships:
 $\langle\hat{\xi}_\text{m}^\dagger(t)\hat{\xi}_\text{m}(0)\rangle=n_\text{m}^\text{i}\delta(t)$, $\langle\hat{\xi}_\text{m}(t)\hat{\xi}_\text{m}^\dagger(0)\rangle=(n_\text{m}^\text{i}+1)\delta(t)$;
 $\langle\hat{\xi}_\text{e}^\dagger(t)\hat{\xi}_\text{e}(0)\rangle=n_\text{e}\delta(t)$, $\langle\hat{\xi}_\text{e}(t)\hat{\xi}_\text{e}^\dagger(0)\rangle=(n_\text{e}+1)\delta(t)$;
 $\langle\hat{\xi}_0^\dagger(t)\hat{\xi}_0(0)\rangle=n_0\delta(t)$, $\langle\hat{\xi}_0(t)\hat{\xi}_0^\dagger(0)\rangle=(n_0+1)\delta(t)$; 
$n_\text{e}\kappa_\text{e}+n_0\kappa_0=n_\text{c}\kappa$.

We will rewrite this in matrix form, setting 
\begin{align*}
 v(t) := \begin{pmatrix} \hat{a}(t)\\  \hat{a}^{\dagger}(t) \\ \hat{b}(t) \\\hat{b}^{\dagger}(t) \end{pmatrix} &&
 w(t) := \begin{pmatrix} \sum_{j=\text{e},0}\sqrt{\kappa_j}\hat{\xi}_j(t) \\  \sum_{j=\text{e},0}\sqrt{\kappa_j}\hat{\xi}_j(t) \\ 
 \sqrt{\gamma_\text{m}}\hat{\xi}_\text{m}(t)\\\sqrt{\gamma_\text{m}}\hat{\xi}_\text{m}(t)  \end{pmatrix}\\
\end{align*}

and
\begin{align*}
 A:= \begin{pmatrix}
     i\Delta-\frac{\kappa}{2} & 0 & ig & ig \\
     0 & -i\Delta-\frac{\kappa}{2} & -ig & -ig \\
     ig & ig &  -i\omega_\text{m}-\frac{\gamma_\text{m}}{2} & 0\\
     -ig & -ig &  0 & i\omega_\text{m}-\frac{\gamma_\text{m}}{2} 
    \end{pmatrix}\\
\end{align*}
thus we get
\begin{align*}
 v'(t) =Av(t)+w(t).
\end{align*}

Using Fourier transforms $ \mathcal{F}(f(t))=f(\omega)\equiv\int_{-\infty}^\infty \! f(t)e^{i\omega t} \, \mathrm{d}t$ we have
\begin{align*}
-i \omega v(\omega) =Av(\omega)+w(\omega)
\end{align*}
which has the solution
\begin{align*}
v(\omega) = \left(-i\omega I- A\right)^{-1}w(\omega)=B^{-1}w(\omega),
\end{align*}
\begin{align*}
 B&= \begin{pmatrix}
    -i\omega- i\Delta+\frac{\kappa}{2} & 0 &- ig &- ig \\
     0 &-i\omega+ i\Delta+\frac{\kappa}{2} & ig & ig \\
     -ig & -ig & -i\omega+ i\omega_\text{m}+\frac{\gamma_\text{m}}{2} & 0\\
     ig & ig &  0 &-i\omega- i\omega_\text{m}+\frac{\gamma_\text{m}}{2} 
    \end{pmatrix}\\
&\equiv \begin{pmatrix}
   1/\chi_c & 0 &- ig &- ig \\
     0 & 1/\bar\chi_c & ig & ig \\
     -ig & -ig & 1/\chi_\text{m} & 0\\
     ig & ig &  0 & 1/\bar\chi_\text{m}
    \end{pmatrix}.
\end{align*}

For cooling we use $\Delta=-\Omega_\text{m}$ and let $\delta=\omega-\Omega_\text{m}$. Applying the rotating wave approximation, omitting contribution from $\bar\chi_c$, $\bar\chi_\text{m}$, we get
\begin{equation}\label{a}
\hat{a}(\omega)=\frac{\chi_c \sum\sqrt{\kappa_i}\hat{\xi}_i+ig\chi_\text{m}\chi_c\sqrt{\gamma_\text{m}}\hat{\xi}_\text{m}}{1+g^2\chi_c\chi_\text{m}}
\end{equation}
\begin{equation}\label{b}
\hat{b}(\omega)=\frac{\chi_\text{m}\sqrt{\gamma_\text{m}}\hat{\xi}_\text{m}+ig\chi_\text{m}\chi_c \sum\sqrt{\kappa_j}\hat{\xi}_j}{1+g^2\chi_c\chi_\text{m}}.
\end{equation}

From the input-output theory the output field can be expressed as 
\begin{equation}\label{out}
\begin{aligned}
\hat{a}_\text{out}=&\hat{\xi}_{e}-\sqrt{\kappa_\text{e}}\hat{a}\\
=&\left(1-\frac{\chi_c\kappa_\text{e}}{1+g^2\chi_c\chi_\text{m}}\right)\hat{\xi}_\text{e}-\frac{\chi_c\sqrt{\kappa_\text{e} \kappa_0}}{1+g^2\chi_c \chi_\text{m}}\hat{\xi}_0-\frac{i\sqrt{\kappa_\text{e} \gamma_\text{m}}g\chi_c \chi_\text{m}}{1+g^2\chi_c\chi_\text{m}}\hat{\xi}_\text{m}.
\end{aligned}
\end{equation}

Since a cancelling tone is added, the detected field is modified as
\begin{equation}\label{out_c}
\begin{aligned}
\hat{a}_\text{out}=&\hat{\xi}_{e}-\sqrt{\kappa_\text{e}}\hat{a}-\hat{\xi}_{e}\\
=&\frac{\chi_c\kappa_\text{e}}{1+g^2\chi_c\chi_\text{m}}\hat{\xi}_\text{e}-\frac{\chi_c\sqrt{\kappa_\text{e} \kappa_0}}{1+g^2\chi_c \chi_\text{m}}\hat{\xi}_0-\frac{i\sqrt{\kappa_\text{e} \gamma_\text{m}}g\chi_c \chi_\text{m}}{1+g^2\chi_c\chi_\text{m}}\hat{\xi}_\text{m}\\
=&\sum_{k=\text{e,0,m}}f_k(\omega)\hat{\xi}_k(\omega).
\end{aligned}
\end{equation}

The spectrum analyser detects the symmetric power spectral density (PSD) \cite{RevModPhys.82.1155}
\begin{equation}\label{spec}
\begin{aligned}
\frac{S(\omega)}{\hbar\omega}=&\frac{1}{2}\int_{-\infty}^{\infty}e^{i\omega t}\langle \hat{a}_\text{out}^\dagger(0)\hat{a}_\text{out}(t)+ \hat a_\text{out}(t)\hat a_\text{out}^\dagger(0)\rangle\mathrm{d}t.\\
=&\frac{1}{2\pi}\frac{1}{2}\langle \hat a_\text{out}^\dagger(-\omega)\hat a_\text{out}(\omega)+ \hat a_\text{out}(\omega)\hat a_\text{out}^\dagger(-\omega)\rangle.
\end{aligned}
\end{equation}
Strictly speaking this is for the lab frame, however in this case only the detunings enter the equations, therefore we could directly substitute in Supplementary Equation~\ref{a} and \ref{out_c} and get
\begin{equation}
\frac{S(\omega)}{\hbar\omega}=\sum_{k=\text{e,0,m}}\vert f_k(\omega)\vert^2\left(n_k^\text{i}+\frac{1}{2}\right),
\end{equation}
$n_{\text{e,0}}^i$ being equivalent to $n_{\text{e,0}}$. Including the added noise from the amplification chain, we get
\begin{equation}
\frac{S(\omega)}{\hbar\omega}=\frac{g^2\kappa_\text{e}\gamma_\text{m}}{\vert g^2+\left(\frac{\kappa}{2}-i\delta\right)\left(\frac{\gamma_\text{m}}{2}-i\delta\right)\vert^2}\left(n_\text{m}^\text{i}+\frac{1}{2}\right)+\frac{\vert\frac{\gamma_\text{m}}{2}-i\delta\vert^2\kappa_\text{e}\kappa}{\vert g^2+\left(\frac{\kappa}{2}-i\delta\right)\left(\frac{\gamma_\text{m}}{2}-i\delta\right)\vert^2}\left(n_\text{c}+\frac{1}{2}\right)+n_\text{add}.
\end{equation}
Note that the noise operators in the frequency domain satisfy the following relations:
 $\langle\hat{\xi}_k^\dagger(\omega)\hat{\xi}_k(\omega')\rangle=2\pi n_k^\text{i}\delta(\omega+\omega')$, $\langle\hat{\xi}_k(\omega)\hat{\xi}_k^\dagger(\omega')\rangle=2\pi(n_k^\text{i}+1)\delta(\omega+\omega')$.

The final occupation can be found via equipartition \cite{PhysRevLett.101.263602, Rocheleau2010}:
\begin{equation}
\begin{aligned}\label{nm}
1+2n_\text{m}=&\frac{\langle\hat{x}^2\rangle}{x_\text{zpf}^2}=\int_{-\infty}^{\infty}\frac{\mathrm{d}\omega}{2\pi}\frac{S_{xx}(\omega)}{x^2_\text{zpf}}\\
=&\frac{1}{x^2_\text{zpf}}\int_{-\infty}^{\infty}\frac{\mathrm{d}\omega}{2\pi}\int_{-\infty}^{\infty}\frac{1}{2}\langle \hat{x}(t)\hat{x}(0)+\hat{x}(0)\hat{x}(t)\rangle e^{i\omega t}\mathrm{d}t,
\end{aligned}
\end{equation}
where $\hat x(t)=x_\text{zpf}(\hat{b}(t)+\hat{b}^\dagger(t))$, $x_\text{zpf}$ being the zero-point fluctuation. Substituting in Supplementary Equation~\ref{b}, using formulae of contour integrals and considering $\kappa^2\gg4g^2,\kappa\gamma_\text{m},\gamma_\text{m}$, we find
\begin{equation}
n_\text{m}\approx\frac{\kappa\gamma_\text{m}}{4g^2+\kappa\gamma_\text{m}}n_\text{m}^\text{i}+\frac{4g^2}{4g^2+\kappa\gamma_\text{m}}n_\text{c}=\frac{1}{C+1}n_\text{m}^\text{i}+\frac{C}{C+1}n_\text{c}.
\end{equation}\\

{\setlength{\parindent}{0cm}
{\bf \large Supplementary Note 7: Noise from the signal generator}
\vspace{1em}

For the measurement shown in Fig.~4 of the main text, we use a Phase Matrix QuickSyn FSW-0020 microwave signal generator. We measure a sideband noise $S_{\phi}$ of $-130$~dBc/Hz at 120~kHz offset for a 5.1 GHz carrier signal. Cavity noise occupancy $n_\text{c}$ contributed by the signal's own noise can be estimated by 

\begin{equation}
n_\text{c} = \frac{P_\text{in}S_{\phi}\cdot\kappa_\text{e}}{\hbar\omega}\frac{\kappa_\text{e}}{(\kappa/2)^2}. 
\end{equation}
For a critically coupled cavity, we have
\begin{equation}
n_\text{c}=\frac{P_\text{in}\times10^{-13}/\text{Hz}}{\hbar\omega}.
\end{equation}
The green line in Fig.~4c in the main text is plotted with the above equation.}

Measurements were also performed using an a Agilent PSG-UNY low phase
noise option microwave signal generator which was available to us for
a short time. Although the phase noise of the PSG-UNY at an offset of
120~kHz is specified to be -137~dBc/Hz for 5~GHz carrier signals, the
total sideband noise (amplitude and phase) was observed to be
-132~dBc/Hz. Although we did not have time to perform a full thermal
calibration of the setup with the PSG-UNY, we observed that the cavity
noise $n_\text{c}$ was about 2 dB lower with the PSG generator with no
evidence of additional mode heating, implying that the final
occupation with the PSG generator would be 4.3 phonons.\\
\clearpage

 {\setlength{\parindent}{0cm}
{\bf \large Supplementary References}